\documentclass[aps,prb,preprint,superscriptaddress,showpacs,showkeys]{revtex4}

\usepackage[dvips]{graphicx}
\usepackage[dvips]{color}
\usepackage{amsmath}

\newcommand{\orb}[3]{#1(#2\textit{#3})}
\newcommand{\bond}[2]{\mbox{#1--#2}}

\begin{document}

\title{Density functional theory study of Fe(II) adsorption and oxidation on goethite surfaces}

\author{Benedict Russell}
\email{bjr27@tcm.phy.cam.ac.uk}
\author{Mike Payne}
\affiliation{Theory of Condensed Matter Group, Cavendish Laboratory,
  University of Cambridge, J J Thomson Avenue, Cambridge CB3 0HE, UK}
\author{Lucio Colombi Ciacchi}
 \altaffiliation{Present address: Hybrid Materials Interfaces Group,
  Faculty of Production Engineering, University of Bremen, 28359 Bremen, Germany}
  \email{colombi@hmi.uni-bremen.de}
\affiliation{Fraunhofer Institut f\"ur Werkstoffmechanik,
  W\"ohlerstrasse 11, 79108 Freiburg, Germany}
\affiliation{Institut f\"ur Zuverl\"assigkeit von Bauteilen und
  Systemen, Universit\"at Karlsruhe, Kaiserstr. 12, 76131 Karlsruhe,
  Germany}

\date{\today}

\begin{abstract}

We study the interactions between Fe(II) aqua-complexes and surfaces
of goethite ($\alpha$-FeOOH) by means of density-functional theory calculations 
including the so-called Hubbard $U$ correction to the exchange-correlation functional.
Using a thermodynamic approach, we find that (110) and (021) surfaces in contact 
with aqueous solutions are almost equally stable, despite the evident needle-like shape 
of goethite crystals indicating substantially different reactivity of the two faces.
We thus suggest that crystal anisotropy may result from different growth rates
due to virtually barrier-less adsorption of hydrated ions on the (021) but not 
on the (110) surface.
No clear evidence is found for spontaneous electron transfer from an 
adsorbed Fe(II) hex-aqua complex to a defect-free goethite substrate.
Crystal defects are thus inferred to play an important role in assisting 
such electron transfer processes observed in a recent experimental study.
Finally, goethite surfaces are observed to enhance the partial oxidation of 
adsorbed aqueous Fe(II) upon reaction with molecular oxygen.
We propose that this catalytic oxidation effect arises from donation of 
electronic charge from the bulk oxide to the oxidizing agent through 
shared hydroxyl ligands anchoring the Fe(II) complexes on the surface.
\end{abstract}



\maketitle

\section{Introduction}

Goethite (\(\alpha\)-FeOOH) is the most common iron (III)
oxyhydroxide, and the only stable phase with respect to hematite
and liquid water at ambient temperature and pressure conditions.\cite{feo-maj-03b}
It occurs naturally in soils as a result of weathering of other iron-based 
minerals, and is the dominant oxyhydroxide phase in lake and 
marine sediments.\cite{feo-zee-03}
Due to its tendency to form nanoscale crystals with high specific surface
area, goethite plays an important role in nature in controlling the mobility
of heavy metals such as As, Cd, Zn, Hg, as well as phosphate anions,
and has been studied extensively as a model adsorbent in environmental science
and technology.\cite{feo-kos-04,feo-way-05,feo-moh-07}
It has also found applications as a precursor in the development of magnetic
recording systems, being converted into maghemite ($\gamma$-Fe$_2$O$_3$)\cite{feo-cor-sch} 
or metallic particles\cite{feo-nun-03} by thermal treatment.
Recently, it has been studied as a model colloidal system showing
magnetically-sensitive liquid crystal behaviour.\cite{feo-thi-07}

In technological applications, it is important both to have a control over the
growth of nanoscale particles (e.g., to produce them with narrow size distributions
and uniform properties), and to be able to predict the interactions between 
the particle surfaces and their external environment (e.g., to optimize
their adsorption capabilities).
However, the microscopic mechanisms responsible for the growth of
goethite crystals from aqueous solutions are still poorly understood,\cite{surf-wei-98}
and detailed electronic structure knowledge of iron oxyhydroxide surfaces
is limited.\cite{surf-wil-04}
This is partly due to the fact that quantum-mechanical calculations of
iron oxyhydroxide phases are challenging for standard methods such as
density functional theory (DFT),\cite{tec-hoh-64,tec-koh-65}
due to the complex magnetic structure and the large crystal unit cells
which these phases present.
Moreover, especially in the case of iron oxides, the strong electronic 
correlations arising from localised \textit{d}-orbitals are not well described 
within the standard local-density (LDA) or generalized gradient (GGA) 
DFT approximations.\cite{ldau-ter-84a,ldau-ter-84b,ldau-dud-98}
In the present paper we undertake an extensive study of the
surface chemistry of goethite within the so-called LDA+$U$ 
scheme\cite{ldau-ani-91b,ldau-ani-93,ldau-sol-94}, which integrates a 
model-Hamiltonian-based treatment of the localised \orb{Fe}{3}{d}
orbitals within the framework of a GGA-DFT calculation.
Our aim is to elucidate fundamental features of the mechanisms of 
goethite surface reactivity, in the context of crystal growth upon
interaction with dissolved iron complexes.

Synthetic goethite may be produced either by precipitation from a saturated 
aqueous Fe(III) solution or by slow oxidation of aqueous Fe(II).\cite{feo-zee-03}
It is well established that the oxidation of Fe(II) to Fe(III) in aqueous solution may 
be self-catalysed by already present iron oxyhydroxide particles, resulting in the growth 
of an Fe(III) layer on the oxide surface that is similar in structure to the underlying bulk 
material.\cite{surf-wei-98,surf-wil-04}
However, the mechanisms of the surface redox reactions remain elusive.
Recently, Williams and Scherer \cite{surf-wil-04} used
M\"ossbauer spectroscopy to study the reaction of aqueous Fe(II) with
Fe(III) oxide surfaces.
Their study gives evidence of spontaneous electron transfer from the adsorbed 
Fe(II) complex to the underlying oxide, most probably occurring via overlap 
of \orb{Fe}{3}{d} orbitals in octahedral edge-sharing environments on the 
crystal surface.
However, whether this is an essential step in the catalytic oxidation of Fe(II)
at the oxide surface, or whether it is in fact a competing process, was left as an 
unanswered question.
Here we attempt to address this issue by performing electronic structure
calculations of Fe(II) complexes adsorbing on goethite surfaces with
different crystallographic orientations.

\begin{figure*}
\includegraphics[width=17.78cm]{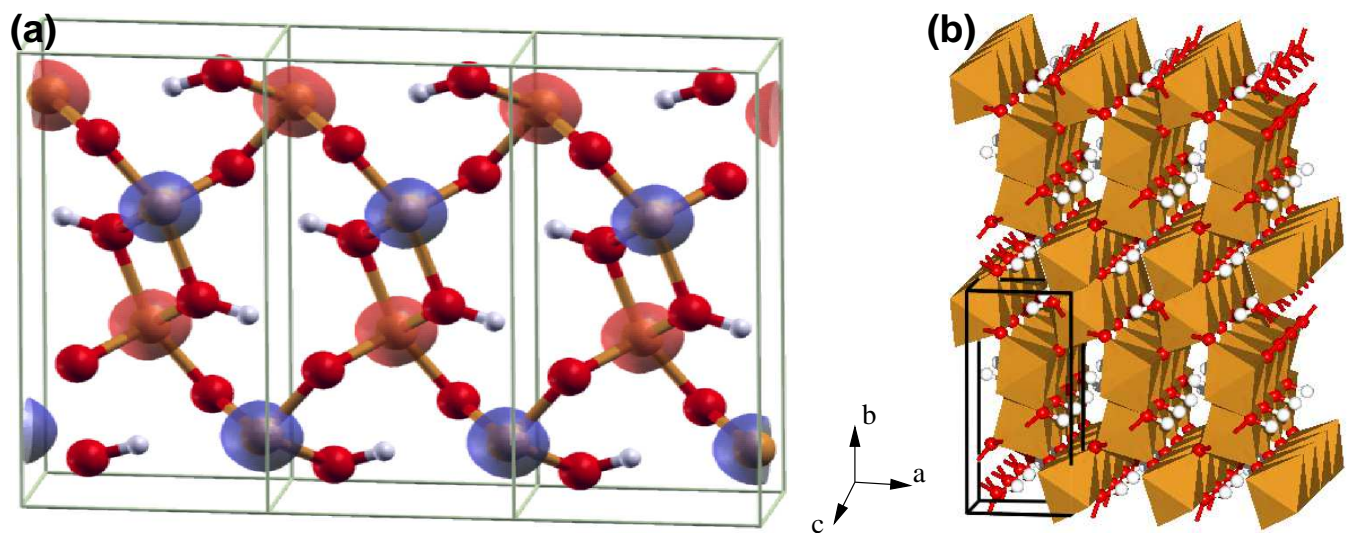}
\caption{(Color online) Two views of the structure of bulk goethite, showing the unit
  cell: (a) with spin density isosurface to show the antiferromagnetic
  arrangement of local spin moments in the ground state; (Red and blue
  indicate up and down spin, respectively.) (b) a polyhedral representation,
  illustrating the double chain structure of octahedrally coordinated Fe
  ions. Colors (online only): Fe gold, O red, H white.\label{fig_goethite_struc}}
\end{figure*}

The crystal structure of goethite has been studied extensively by X-ray and neutron
diffraction, and is shown in fig.~\ref{fig_goethite_struc}.
%
%
Goethite crystallises with an orthorhombic unit cell, with symmetry usually 
described by the space group \textit{Pbnm},\cite{feo-for-68} though 
the orthorhombic group \textit{Pnma} may be used equivalently.\cite{feo-szy-68}
The unit cell contains four FeOOH formula units, with ionic positions given 
by $\pm(x,y,\frac{1}{4})$ and $\pm(\frac{1}{2}+x,\frac{1}{2}-y,\frac{3}{4})$.
The structure of goethite may be described as a slightly distorted hexagonally 
close-packed arrangement of oxygen and hydroxyl anions along the cell [100] axis, 
with Fe cations occupying half of the octahedral interstitial sites.
In a polyhedral representation, it consists of parallel double chains of
edge-linked FeO$_3$(OH)$_3$ octahedra along the [001] direction, with
neighboring chains linked to each other by corner-sharing.
Below the N\'{e}el temperature of approximately 400K, goethite is
antiferromagnetic, with local magnetic moments on the Fe ions
alternating along the cell \(b\)-axis, and with the moments aligned
parallel to the cell \(c\)-axis.\cite{feo-for-68,feo-szy-68}
Goethite is generally considered to be a charge-transfer insulator with
a band gap of about 2.5 eV, the top of the valence band being dominated
by \orb{O}{2}{p} states and the bottom of the conduction band 
having predominantly \orb{Fe}{3}{d} character.\cite{feo-she-05}

\begin{figure*}
\includegraphics[width=17.78cm]{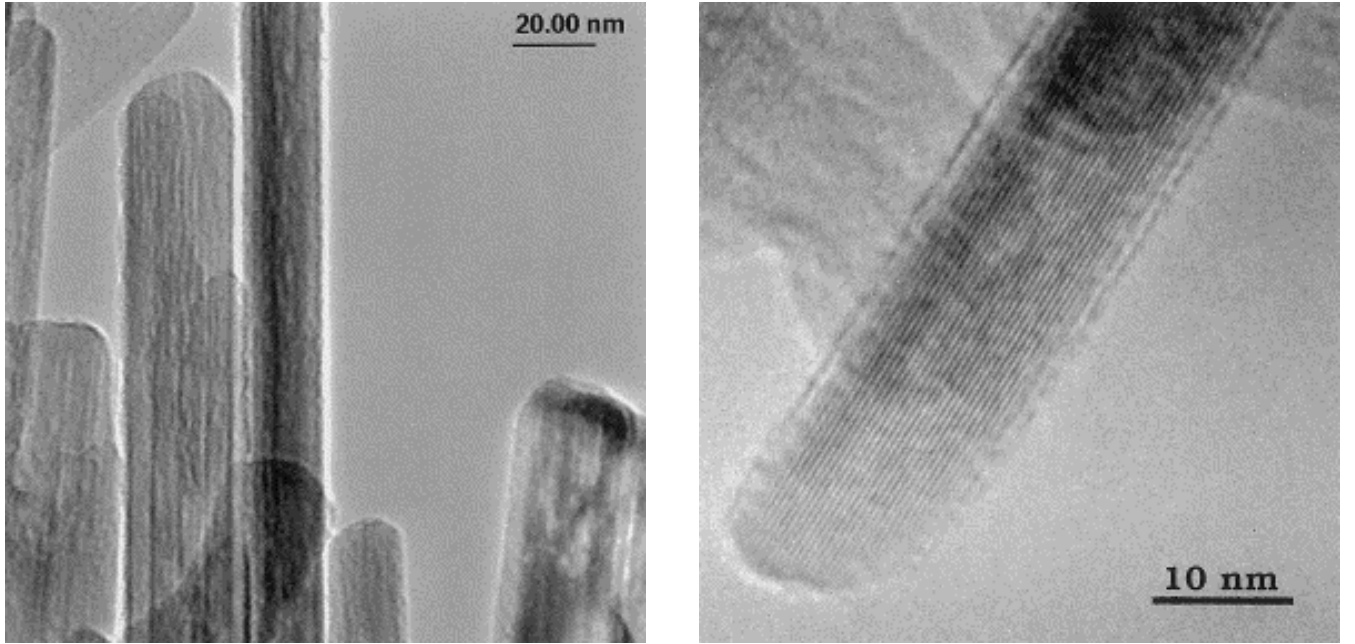}
\caption{TEM images of synthetic goethite crystals. Reprinted with 
permission from Elsevier.\cite{surf-boi-01}\label{fig_goethite_TEM}}
\end{figure*}

Natural and synthetic goethite crystals present a needle-like morphology,
as illustrated by the transmission electron microscopy images in
fig.~\ref{fig_goethite_TEM}. 
%
%
The crystal surface is usually made up mostly of (110) faces running parallel to the long axis 
of the needles, while the ends of the needles are capped predominantly by (021) and other planes
with a large component perpendicular to the cell $c$-axis.%
\cite{surf-smi-83,surf-sch-84,surf-amo-86,surf-smi-87,surf-ran-99}
The \{110\} and \{021\} families might therefore represent surfaces with
distinctly different character: the former being relatively stable and the latter 
providing a site for active crystal growth.
In the remainder of the paper, we will focus on these two surfaces.
In section \ref{goethite} we will study their structure and relative stability.
In section \ref{autocat} we will then look for evidence of spontaneous
oxidation of Fe(II) hex-aqua complexes adsorbing on them.
Finally, in section \ref{sec:oxi_mol_oxy} we will investigate the possibility
of surface-catalyzed oxidation via reaction of the adsorbed complexes with 
molecular oxygen.

\section{Methods}\label{methods}

\subsection{Density Functional Calculations}

The calculations described in this paper have been performed using the
{\sc Castep} simulation package,\cite{tec-cla-05} which provides an
implementation of spin-polarized DFT based on the plane-wave
pseudopotential scheme.\cite{tec-pay-92}
Exchange and correlation were treated within the generalized gradient
approximation (GGA), using the functional form of Perdew, Burke and
Enzerhof (PBE).\cite{tec-per-96}
Ultrasoft pseudopotentials\cite{tec-van-90,tec-laa-93} were used to
describe the ionic cores ($1s^2$ for O, $1s^22s^22p^63s^23p^6$ for Fe),
and non-linear core corrections\cite{tec-lou-82} were applied
for iron to improve the description of magnetically polarized states.
The electronic Kohn-Sham wave functions were expanded using a plane-wave basis set,
up to a kinetic energy cut-off of 450~eV, which was shown to converge
the formation energy of bulk goethite to within 1~meV per
atom with respect to increasing basis size.
Monkhorst-Pack grids\cite{tec-mon-76} were used to sample the Brillouin zone.
In the case of bulk goethite, a $4 \times 2 \times 6$ grid was used, giving
convergence of better than 0.1~meV in the total energy per formula unit.

The LDA+$U$ formalism has been implemented within {\sc Castep} according
to the scheme described by Cococcioni et
al..\cite{ldau-coc-02,ldau-coc-05}
Within this scheme, the Hubbard $U$ is not treated as an empirical fitting parameter,
but may rather be determined self-consistently from the calculated ground state.
In principle, the value of $U$ should be determined separately for
each system studied.
However, it would be inappropriate to make direct comparisons between
total energies from calculations using differing values of $U$.
For this reason, we have instead determined the self consistent value
for Fe in bulk goethite, and then used this same value for the
subsequent surface calculations.
The possible impact of using a different value for the $U$ parameter is considered briefly
in section~\ref{autocat_021}.

\subsection{Thermodynamic Approach}\label{comp_thermo}

By means of DFT calculations, we can readily obtain total energies at
zero temperature for a range of surface configurations. However, in
order to make a more meaningful comparison of the relative stability
at finite temperature of surface structures of different
stoichiometry, we take a thermodynamical approach, which we outline below.
A more detailed discussion relating to the
integration of thermodynamics with electronic structure calculations
may be found, for example, in Ref.~\onlinecite{surf-reu-01}.

Conceptually, we consider a system consisting of three regions: a
large region of bulk goethite with stoichiometry FeOOH, a large
reservoir of liquid water at neutral pH and a surface region of variable
stoichiometry.
The most stable surface configuration at given
temperature, $T$, and pressure, $p$, is that which
minimizes the surface free energy, $\gamma(T,p)$, given, in the case
of a slab model with two equivalent surfaces, by
\begin{widetext}
\begin{equation}\label{thermo_gamma}
\gamma(T,p) = \frac{1}{2A}
\left[G_{slab}(T,p,N_{\mathrm{Fe}},N_{\mathrm{O}},N_{\mathrm{H}})
-N_{\mathrm{Fe}}\mu_{\mathrm{Fe}}(T,p)
-N_{\mathrm{O}}\mu_{\mathrm{O}}(T,p)
-N_{\mathrm{H}}\mu_{\mathrm{H}}(T,p)
\right],
\end{equation}
\end{widetext}
where $\mu_{\mathrm{Fe}}$, $\mu_{\mathrm{O}}$, $\mu_{\mathrm{H}}$
are the chemical potentials for Fe, O and H atoms, respectively, 
$N_{\mathrm{Fe}}$, $N_{\mathrm{O}}$, $N_{\mathrm{H}}$ are the numbers
of atoms of each species making up the surface region, and
$A$ is the area of the surface unit cell.

At chemical equilibrium, we can impose relationships between the
chemical potentials of Fe, O and H:
\begin{eqnarray}
\mu_{\mathrm{Fe}}+2\mu_{\mathrm{O}}+\mu_{\mathrm{H}}&=g^{bulk}_{\mathrm{FeOOH}},\\
2\mu_{\mathrm{H}}+\mu_{\mathrm{O}}&=g^{liquid}_{\mathrm{H}_2\mathrm{O}},
\end{eqnarray}
where $g$ represents a Gibbs free energy per formula unit.
All the surface configurations considered in this paper may be
constructed stoichiometrically from FeOOH and H$_2$O, and hence we can
rewrite eq.~\eqref{thermo_gamma} as:
\begin{widetext}
\begin{equation}\label{thermo_gamma_2}
\gamma(T,p) = \frac{1}{2A}
\left[G_{slab}(T,p,N_{\mathrm{FeOOH}},N_{\mathrm{H}_2\mathrm{O}})
-N^{\,}_{\mathrm{FeOOH}}\,g^{bulk}_{\mathrm{FeOOH}}(T,p)
-N^{\,}_{\mathrm{H}_2\mathrm{O}}\,g^{liquid}_{\mathrm{H}_2\mathrm{O}}(T,p)
\right].
\end{equation}
\end{widetext}

The Gibbs free energies, $G_{slab}$, $g^{bulk}_{\mathrm{FeOOH}}$ and
$g^{liquid}_{\mathrm{H}_2\mathrm{O}}$, may be related to the
total energy, $E^{el}$, obtained in a typical DFT calculation, by 
\begin{equation}
G = E^{el}+E^{vib}+pV-TS
\end{equation}
For pressures of the order of 1 atm, and surface relaxations of the order
of 1 \AA, the contribution from the $pV$ term to $\gamma$ is of
the order of 0.001 meV/\AA$^2$, and may therefore 
safely be neglected.
For solid materials, the remaining terms, $E^{vib}-TS$, arise
principally from lattice vibrations (phonons).
In this work, we assume that the phonon density of states of the
solid is not significantly altered by the surface configuration, so
that contributions from the slab and from the bulk will cancel out to
a large extent in determining the surface free energy.
Hence, for the purpose of comparing surface free energies of different 
faces, we take the DFT total energies of the slab and the bulk as a direct
approximation to the corresponding Gibbs free energies.

In order to estimate $g^{liquid}_{\mathrm{H}_2\mathrm{O}}(T,p)$, we combine a
well converged DFT total energy for a single water molecule,
$E^{el}_{\mathrm{H}_2\mathrm{O}}$, with an experimental Gibbs free
energy of solvation, $\Delta g^{solv}_{\mathrm{H}_2\mathrm{O}}$, so that:
\begin{equation}
g^{liquid}_{\mathrm{H}_2\mathrm{O}}(T,p)=
E^{el}_{\mathrm{H}_2\mathrm{O}}+
\Delta g^{solv}_{\mathrm{H}_2\mathrm{O}}(T,p).
\end{equation}

In an aqueous environment, we should also consider the
free energy of solvation of the surface, $\Delta
G^{\vphantom{q}solv}_{slab}$. 
However, the chosen surface terminations represent in some
sense an explicit consideration of the first stages of
hydration of the bare surfaces, and the fully hydroxylated (110) and (021) surfaces 
have a very similar density of surface anion groups (approximately
15~nm$^{-2}$ in both cases). 
Thus, even if the solvation free energy is a significant
fraction of the overall surface energy, it is unlikely to contribute
significantly to the relative stability of these surfaces.
For this reason, no attempt has been made to include $\Delta
G^{\vphantom{q}solv}_{slab}$ in the results presented in this paper.

We thus arrive at our final expression:
\begin{widetext}
\begin{equation}\label{thermo_gamma_approx}
\gamma(T,p) \simeq \frac{1}{2A}
\left[E^{el}_{slab}
-N^{\,}_{\mathrm{FeOOH}}\,E^{el}_{\mathrm{FeOOH}}
-N^{\,}_{\mathrm{H}_2\mathrm{O}}\,\left(E^{el}_{\mathrm{H}_2\mathrm{O}}+
\Delta g^{solv}_{\mathrm{H}_2\mathrm{O}}(T,p)\right)
\right].
\end{equation}
\end{widetext}
In the results that follow, we use a value $\Delta
g^{solv}_{\mathrm{H}_2\mathrm{O}}(298\,\mathrm{K},1\,\mathrm{atm})=-0.274~\mathrm{eV/molecule}$,
taken from Ref.~\onlinecite{oth-ben-84}.
The value of
$E^{el}_{\mathrm{FeOOH}}$ is taken from the calculations described in section
\ref{goethite_bulk}, while $E^{el}_{\mathrm{H}_2\mathrm{O}}$ is
obtained from a geometry optimization of an isolated water molecule in
a 16~\AA\ cubic supercell.
In both cases, the same cut-off energy, pseudopotentials and
exchange-correlation functional were used as for the main calculations.

\section{Goethite Surfaces}\label{goethite}

In this section we will present results on the thermodynamic
stability of goethite surfaces in equilibrium with water
solutions.
After a brief description of the bulk properties of goethite
crystals, structural and energetic details of surfaces
will be described and discussed in the context of crystal
growth from dissolved iron ions.

\subsection{Bulk Goethite}\label{goethite_bulk}

In our GGA-DFT calculations, we have taken into account five possible
magnetic phases of goethite: a non-magnetic (NM) phase, a
ferromagnetic (FM) phase and three antiferromagnetic (AFM) phases
differing in the ordering of up and down local spin moments within the
cell.
The cell parameters and energies resulting from structural
optimizations of each phase are shown in Table~\ref{table_goethite_bulk}. 

%
%
\begin{table}
\caption{Calculated and average experimental cell parameters, 
magnetic moments and relative energy per formula unit for 
different magnetic structures of goethite.
\label{table_goethite_bulk}}
\begin{center}
\begin{tabular}{ll|lll|l}
\hline
\hline
&&$a$(\AA)&$b$(\AA)&$c$(\AA)&$\Delta E$ (meV)\\
\hline
Exp.&AF&4.625&9.963&3.023&~\\
\hline
GGA&AF&4.660&9.987&3.006&0\\
&FM&4.402&9.594&2.898&+100\\
&NM&4.369&9.514&2.908&+155\\
\hline
GGA+$U$&AF&4.646&10.150&3.084&~\\
\hline
\hline
\end{tabular}
\end{center}
\end{table}

In agreement with experimental results,\cite{feo-for-68,feo-szy-68}
we find an AFM ground state with local spin moments of the Fe ions
alternating along the cell $b$-axis.
The optimized cell parameters and atomic coordinates agree to within
1\,\% and 0.3\,\%, respectively, with the experimental values
(Tables~\ref{table_goethite_bulk} and~\ref{table_goethite_internal}).
%
%
\begin{table}
\caption{Calculated and experimental ionic positions for
  antiferromagnetic goethite. Experimental values from
  Ref.~\onlinecite{feo-for-68}}
\label{table_goethite_internal}
\begin{center}
\begin{tabular}{l|ll|ll|ll}
\hline
\hline
~&\multicolumn{2}{c}{Experimental}&\multicolumn{2}{c}{GGA}&\multicolumn{2}{c}{GGA+$U$}\\
\cline{2-7}
Species&\(x\)&\(y\)&\(x\)&\(y\)&\(x\)&\(y\)\\
\hline
Fe&0.0477&0.8539&0.0465&0.8509&0.0633&0.8559\\
O\(_I\)&0.7058&0.1994&0.7039&0.2015&0.6775&0.1937\\
O\(_{II}\)&0.1974&0.0531&0.1965&0.0529&0.1871&0.0585\\
H&0.3991&0.0876&0.3982&0.0847&0.3899&0.0947\\
\hline
\hline
\end{tabular}
\end{center}
\label{goethite-xy}
\end{table}
By varying the cell volume around the equilibrium value and fitting a 
Murnaghan equation of state to the resulting total energies, the bulk 
modulus is estimated to be 89 GPa. 
As is often found in GGA calculations, this is considerably lower than
the recently reported experimental value of 111~GPa.\cite{feo-nag-03}

For the minimum energy structure, the electronic density of states (DOS) 
projected onto \orb{Fe}{3}{d} and \orb{O}{2}{p} atomic orbitals 
(Fig.~\ref{fig_bulk_110_DOS}) reveals the presence of a band gap of 0.8~eV,
which is significantly smaller than the experimental value of 2.5 eV. 
We also observe strong mixing of the majority-spin \orb{Fe}{3}{d} 
states with the \orb{O}{2}{p} states across the whole range of the 
valence band, in conflict with the experimental observation that 
the valence band edge should be dominated by \orb{O}{2}{p} states. 
The predominantly \orb{Fe}{3}{d} nature of the conduction band 
is correctly reproduced.
Hybridization of the atomic orbitals in the crystal environment
results in a non-negligible contribution to the density of states
below the Fermi level from minority spin \orb{Fe}{3}{d}-like states.
However, there is no sharp peak below the Fermi level
corresponding to a single occupied minority spin \orb{Fe}{3}{d} 
orbital, as typically found in the case of of Fe$^{2+}$ ions.
This fact, together with the complete occupation of the majority 
spin \orb{Fe}{3}{d} orbitals, is consistent with the identification 
of goethite as an Fe(III) compound.

In order to improve the description of the electronic properties,
we have performed additional calculations for the AFM ground state at the
GGA+$U$ level.
The value of the Hubbard $U$ parameter was determined
self-consistently according to the procedure of Cococcioni et
al.,\cite{ldau-coc-05,ldau-kul-06} giving a value of
$U_{\text{scf}}=5.2$~eV.
The new equilibrium lattice parameters (see Table~\ref{table_goethite_bulk})
are on average slightly further from the most recent experimental
results than the GGA parameters, but the agreement, within 1.5\,\%, is
still reasonable.
With the $U$ correction, the calculated bulk modulus is 109~GPa,
very close to the experimental value.
The optimized ionic positions (see Table~\ref{table_goethite_internal})
present a maximum deviation of 3\,\%\ (average 1\%) 
from the experimental values. 
We note that our self-consistent value of $U$ is implicitly chosen to
correct the electronic properties rather than the structural
properties, which were in any case well described at the GGA level.
Indeed, as reported previously by other authors, it is often not
possible to choose a single value of $U$ that gives quantitatively
correct predictions of both the structural and electronic properties.\cite{ldau-roh-03}

With this in mind, we have calculated the GGA+$U$ electronic DOS 
using the relaxed structure obtained from the GGA calculations.
The results, projected as before onto \orb{Fe}{3}{d} and \orb{O}{2}{p}
orbitals, are shown in fig.~\ref{fig_bulk_110_DOS}b.
The addition of the on-site repulsion term to the Kohn-Sham
Hamiltonian has a strong influence on the electronic structure around
the Fermi level compared with the GGA results.
The unoccupied \orb{Fe}{3}{d} minority spin states are pushed
to higher energies, thus increasing the band gap from the GGA value of
0.8 eV to 2.5 eV.
At the same time, the occupied \orb{Fe}{3}{d} majority spin states
move further below the Fermi level, breaking the strong
\mbox{\orb{Fe}{3}{d}--\orb{O}{2}{p}} hybridization observed in the GGA
DOS.
Thus, GGA+$U$ predicts for Goethite, in full agreement with
experiment, a charge transfer insulating state with a band gap of
$2.5$ eV between \orb{O}{2}{p}-dominated states at the top of the
valence band and localized \orb{Fe}{3}{d} minority spin states at the
bottom of the conduction band.

The GGA+$U$ DOS also shows a reduction in the partial occupation of
minority spin \orb{Fe}{3}{d} states relative to the GGA electronic
structure, making the identification of the Fe(III) oxidation state
even clearer than before.
This also leads to an increase in the local magnetic
moment on the Fe ions from 3.74 $\mu_B$ to 4.16 $\mu_B$.
(To the best of our knowledge, no experimental information
on the value of the magnetic moments is available for
well-crystallized goethite, although Bocquet and Kennedy found a
saturation magnetic moment of 3.80 $\mu_B$ per iron atom for
fine particle goethite.\cite{feo-boc-92})

\subsection{Goethite surfaces}\label{surf}

\subsubsection{Surface terminations}

Previous theoretical studies of goethite surfaces have
assumed complete hydroxylation of the surface.\cite{surf-rus-96b}
Fully hydroxylated surfaces can be considered as arising from a
truncation of the bulk such that all Fe ions remain octahedrally
coordinated, followed by the addition of sufficient protons to cancel
the excess negative charge of the surface layer.
In this study, we take a slightly different approach which allows us
also to investigate a range of intermediate stages of hydroxylation.
Starting from stoichiometric, non-polar terminations of the (110) and
(021) surfaces, we consider the results of heterolytic dissociation
and adsorption of water molecules, obtaining a total of five different
surface configurations for the two surfaces, as outlined below.
Here we implicitly assumed a solution at neutral pH, so that only whole 
water molecules (undissociated or dissociated) are used to terminate the
surfaces.

\begin{figure*}
\includegraphics[width=16cm]{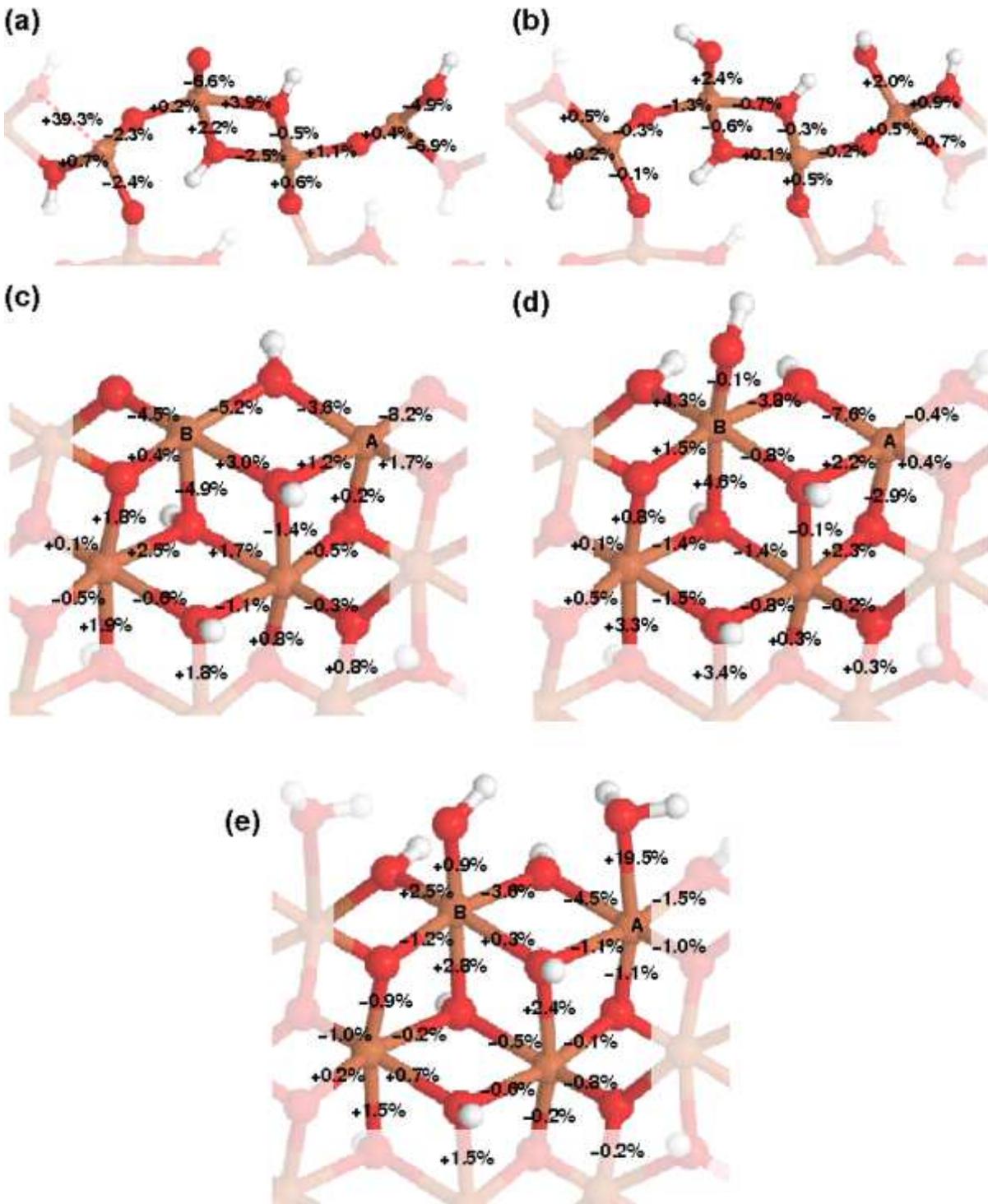}
\caption{(Color online) Changes in Fe---O bond lengths relative to
  corresponding values in the bulk for the (a) bare and
  (b) hydroxylated (110) surfaces and (c) bare, (d) partially 
  hydroxylated and (e) fully hydroxylated (021) surfaces of 
  goethite, as determined by GGA geometry optimizations.\label{fig_surf_bonds}}
\end{figure*}

For the (110) surface, a stoichiometric truncation of the bulk
(fig.~\ref{fig_surf_bonds}a) leads to two inequivalent Fe surface
sites, one five-fold coordinated and the other six-fold coordinated.
%
%
Adjacent six-fold coordinated sites are joined by bridging oxygen 
atoms. 
The dissociative addition of a single water molecule is then
sufficient to fully hydroxylate the surface, with the hydroxyl group
binding to the formerly under-coordinated Fe and the proton donated to the
bridging oxygen atom (fig.~\ref{fig_surf_bonds}b).
The bare (021) surface (fig.~\ref{fig_surf_bonds}c) also presents two
inequivalent Fe sites, in this case both five-fold coordinated. 
One (A) is bound to 3 oxygens and 2 hydroxyl groups, while the other
(B) is bound to 2 oxygens and 3 hydroxyl groups.
The surface anion layer takes the form of bridging oxygen atoms and 
hydroxyl groups. 
We consider two steps of hydroxylation: first, the dissociative addition
of a single water molecule per two Fe sites, with the hydroxyl group binding
to the B site and the proton donated to the bridging oxygen atom
(fig.~\ref{fig_surf_bonds}d), and second, the adsorption of an
additional water molecule onto the Fe A site (fig.~\ref{fig_surf_bonds}e).

\subsubsection{Surface structure}

Starting from the GGA-relaxed goethite bulk structure, the geometries 
of all five surfaces were optimized at the GGA level. 
Using the optimized geometries, single point electronic minimizations were
also performed at the GGA+$U$ level, using the same value of $U=5.2$~eV 
for the \orb{Fe}{3}{d} orbitals as for the goethite bulk. 
In all cases, a slab model was employed, with neighboring slabs
separated from each other by a vacuum region of at least 5.5 \AA. 
Doubling the separation between the slabs was shown to change the resulting
surface energies by less than 1~meV/\AA$^2$.
For both the (110) and the (021) surfaces, the slab contained four 
layers of Fe ions, while the cell vectors in the plane of the surface
corresponded to a single unit cell of the bulk structure.
Brillouin zone sampling was performed by means of Monkhorst-Pack grids in the
plane of the surface, using $8 \times 2$ points for the (110) surfaces and $4
\times 2$ for the (021) surfaces.
The relaxed structures for the three (021) and the two (110) surfaces
are shown in Fig.~\ref{fig_surf_bonds} along with the differences in 
bond lengths with respect to bulk goethite.

For the bare (021) surface, the under-coordination results in a 
reduction of the Fe-O distances.
In the bare (110) surface, beside a reduction of the Fe-O distances,
more drastic relaxations are observed.
The under-coordinated Fe ions at the surface are pulled in towards
the bulk.
In the subsequent rearrangement, formerly three-fold coordinated 
hydroxyl groups break one of their \bond{Fe}{O} bonds to
adopt a bridging configuration between neighboring five-fold
coordinated Fe ions.

In the partially hydroxylated (021) surface, the protonation of the
bridging oxide groups causes an increase in the corresponding Fe--O
bond lengths, while the bridging OH groups are displaced from their
symmetrical positions, with new bond lengths of 2.03 and
\mbox{1.97~\AA}\ to the six-fold and five-fold coordinated Fe ions,
respectively.
The newly added terminal OH group has a bond length of \mbox{1.97
\AA}\ to the surface.
In the fully hydroxylated (021) surface, the additional water ligands
are only weakly bound to the surface, with a Fe--O distance of
2.53~\AA.
However, this is sufficient to satisfy the under-coordination
of the surface Fe ions.
The bridging anions thus return to symmetrical positions, while the 
terminal OH group moves away slightly from the surface, with a 
new \bond{Fe}{O} bond length of 1.99~\AA.
As expected, the hydroxylated (110) surface shows the smallest
relaxations among all five surfaces, with only the outermost anion 
layer showing bond length changes greater than 1\% relative to the bulk.
This justifies the use of a relatively thin slab for the study of
this surface. 

\subsubsection{Surface stability and crystal growth}
\label{subsec:surf_stabil}

The surface Gibbs free energies of each of the five surfaces, calculated
relative to bulk goethite and liquid water as described in
Section~\ref{comp_thermo}, are reported in Table~\ref{table_gamma}.
%
%
\begin{table}
\caption{Calculated surface free energies, $\gamma$, for (110) and (021) surfaces
  of goethite with a number of different surface terminations. The
  geometry of all surfaces was optimized at the GGA level
  only. $U=5.2$~eV for GGA+$U$ calculations.
\label{table_gamma}}
\begin{center}
\begin{tabular}{l|cc}
\hline
\hline
~&\multicolumn{2}{c}{$\gamma$ (meV/\AA$^2$)}\\
~&GGA&GGA+$U$\\
\hline
(110) bare&49&67\\
(110) hydroxylated&21&14\\
\hline
(021) bare&34&46\\
(021) partially hydroxylated&20&14\\
(021) fully hydroxylated&16&12\\
\hline
\hline
\end{tabular}
\end{center}
\end{table}
It is clear that the hydroxylated surfaces are in general strongly favored 
over the bare surfaces both at the GGA and at the GGA+U levels,
with surface energy values between 10 and 20~meV/\AA$^2$.
These are relatively small compared with typical values calculated for
other iron oxide phases.
For example, for all trivial surface terminations of the hydroxylated
hematite(0001) surface Trainor\textit{et al.\@} computed surface free
energies greater than 50~meV/\AA$^2$ at low oxygen partial
pressure.\cite{surf-tra-04}
For the same surface in a dry environment, Wang \textit{et al.\@}
computed a surface energy above 90~meV/\AA$^2$ in all cases except in
very oxygen-rich conditions, in which case a minimum surface energy of
45~meV/\AA$^2$ was observed for the oxygen-terminated surface.\cite{surf-wan-98}
Finally, similar differences in surface energy between the dry or
hydroxylated surfaces of goethite and hematite have been recently
calculated within the Born model of solids.\cite{surf-del-07}
These results may be consistent with the observation that goethite is
commonly formed as the first precipitate from solution even when
hematite is the thermodynamically stable bulk phase.
Indeed, in the initial nucleation stages a phase with very low surface
energy could be favored over a phase with a surface energy too large
to be compensated for by the energy gained from forming the
bulk material.

It is intriguing, given the evident anisotropy of goethite crystals
(see Fig.~\ref{fig_goethite_TEM}), that the (110) and (021) surfaces
present roughly the same surface energy.
Unless very different oxygen terminations that those considered here govern 
the behaviour of the experimentally investigated systems, our finding suggests 
that the needle-like crystal shape may result not from unequal thermodynamical 
surface stabilities but from unequal growth rates, with the (021) surface growing 
outwards faster than the (110) surface.\cite{surf-wei-98}
In an attempt to rationalize this hypothesis, we note that 
our calculated energies of the partially and fully hydroxylated (021) surfaces
differ by less than 4 meV/\AA$^2$.
This corresponds to a binding energy per added water molecule of just
$-0.10$ eV, less than half the energy of a typical single hydrogen
bond in liquid water.\cite{oth-sur-00}
Therefore, the terminal water positions may be only partially
occupied when averaged over time, giving the possibility
of nearly barrier-less adsorption of additional Fe$^{3+}$ 
or Fe$^{2+}$  ions  at these sites.
In contrast, the corresponding sites on the (110) surface are occupied 
by stably bound hydroxyl ligands.
Some form of ligand exchange would thus be required for an additional
ion to bind to this surface, which would introduce an associated energy
barrier.

Notably, goethite crystals take approximately the same form regardless
of whether they are grown by precipitation from an Fe(III)
solution~\cite{surf-boi-01} or by oxidation of an Fe(II)
solution.\cite{surf-wei-98}
However, in the latter case another possible contribution to unequal 
growth rates would be preferential oxidation of Fe(II) to Fe(III) on 
(021) surfaces as compared with (110) surfaces. 
The oxidation of Fe(II) at these goethite surfaces is addressed in the
remaining sections of this paper.

\section{Spontaneous Oxidation upon Adsorption}\label{autocat}

Fe(III) oxide surfaces are thought to promote the autocatalytic
oxidation of Fe(II) ions during crystal growth.\cite{surf-wei-98}
In this section, we aim to investigate the adsorption of a Fe(II)
complex on the (110) and (021) goethite surfaces, in particular
looking for possible spontaneous electron transfer processes to the
Fe(III) solid from the adsorbed ion.

Owing to the large computational cost associated with our
first-principles approach, an exhaustive search for the lowest energy
conformation of an Fe(II) ions adsorbing on each surface is not
presently feasible.
However, it has been observed experimentally that the binding of
octahedral metal complexes on iron oxyhydroxide surfaces strongly
favors adsorption positions which maintain the underlying anionic
stacking sequence.\cite{surf-man-00}
Similarly, experiments have shown that the oxide layer formed when
aqueous Fe(II) ions are adsorbed and oxidized on a Fe(III)
oxyhydroxide surface is generally similar in structure to the
underlying bulk oxyhydroxide.\cite{surf-wil-04}
As explained here below, applying these considerations allows us
to consider only one adsorbed configuration on each of the two 
surfaces, which can both be considered to be fully hydroxylated 
in light of the surface energies computed in the previous section.

\begin{figure}
\includegraphics[width=8.6cm]{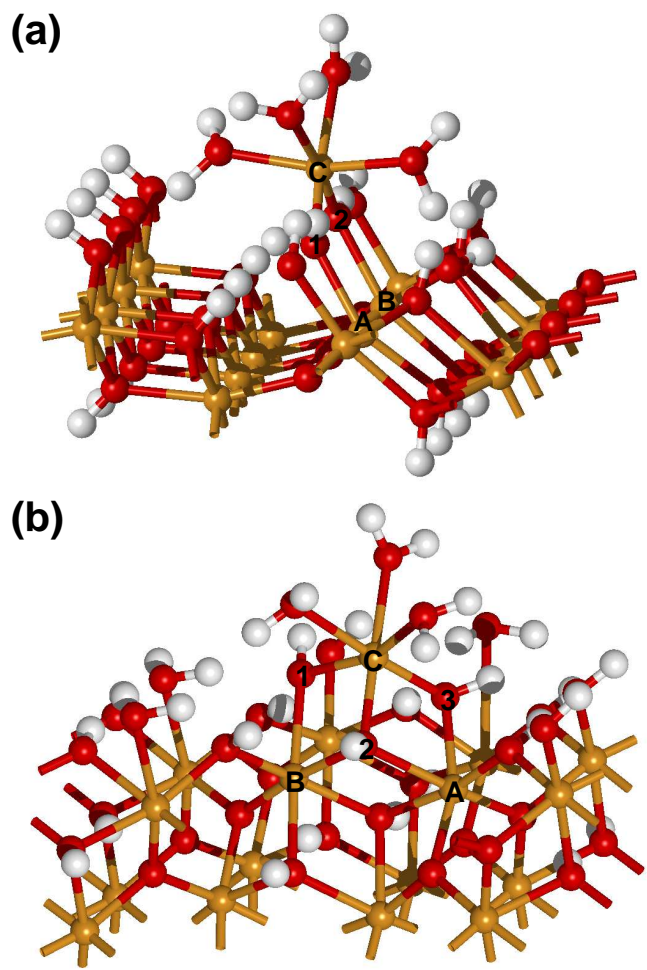}
\caption
{(Color online) The GGA relaxed structure of [Fe(H$_2$O)$_6$]$^{2+}$ bound 
  to the goethite (110) surface in a double corner sharing
  configuration (a) and to the goethite (021) surface in a 
  double edge sharing configuration (b).\label{fig_110_021_complex}}
\end{figure}

On the (110) surface, the complex may bind either through a single
bridging hydroxyl group on the surface or through two terminal
hydroxyl groups, corresponding to single or double corner sharing,
respectively, in terms of the Fe(O,OH,H$_2$O)$_6$ coordination
octahedra.
Both sites may be filled independently, but we may reasonably expect
a stronger interaction between the complex and the surface 
in the double corner sharing case.
The relaxed structure for this site is shown in
fig.~\ref{fig_110_021_complex}a.
%
%
The (021) surface, as previously described, represents a termination
of the double chains of octahedra composing the bulk.
We thus consider the addition of a new octahedron to the end of a
double chain, respecting the intrinsic staggered stacking sequence.
This leads to the adsorbed ion configuration shown in
fig.~\ref{fig_110_021_complex}b, in which two octahedral edges
(i.e. three anion groups) are shared between the complex and the
surface.
In order to reduce unfavorable steric clashes, a water ligand which
occupied a bridging position between the surface and the complex is
replaced with a hydroxyl group.
While this allows the complex to adopt a less strained position on
the surface, preliminary electronic structure calculations showed that
this deprotonation does not influence the oxidation state of the adsorbed complex.

Simulation cells were constructed by starting from the relaxed
surface structures obtained in section~\ref{surf} and expanding the
unit cell to enable the adsorption of a single complex without
significant interaction between the complex and its periodic images in
neighboring cells.
For the (110) surface, the new cell consisted of four original surface
cells, and had surface dimensions $11.0\times12.0$ \AA$^{2}$.
For the (021) surface, the original cell was doubled to give surface
dimensions $9.4\times11.7$ \AA$^{2}$.
In both cases, the number of k-points in the plane of the surface was
reduced to a $2\times2$ grid. 
The vacuum region was also slightly widened to make room for the 
complex, ensuring a \mbox{5~\AA}~separation from the opposing 
surface in the adjacent cell.
For both surfaces, we have assumed the Fe ion in the bound complex
to be in a high spin state, since both the initial state (the unbound Fe(II) 
complex) and the final state (Fe(III) in bulk goethite) fall into this category.
Thus the total starting spin of the system was set to 4/2, equal to
that of the isolated complex.

The structure of both systems was relaxed using GGA DFT calculations
(fig.~\ref{fig_110_021_complex}). 
Relaxed bond lengths compared with corresponding values in bulk goethite 
are given in table~\ref{table_comp_surf}. 
%
%
\begin{table}
\caption{GGA-optimized bond lengths for [Fe(H$_2$O)$_6$]$^{2+}$ 
bound to the goethite (110) and (021) surfaces, compared with the equivalent 
bonds in bulk goethite. Atoms are labelled as in fig.~\ref{fig_110_021_complex}.
\label{table_comp_surf}}
\begin{center}
\begin{tabular}{l|cc}
\hline
\hline
Bond&\multicolumn{2}{c}{Length (\AA)}\\
~&Optimized&Bulk\\
\hline
(110) surface:&&\\
~Fe$^C$---O$^1$&2.003&1.971\\
~Fe$^C$---O$^2$&1.944&1.971\\
~Fe$^C$---Fe$^A$&3.729&3.427\\
~Fe$^C$---Fe$^B$&3.515&3.427\\
\hline
(021) surface:&&\\
~Fe$^C$---O$^1$&2.018&1.971\\
~Fe$^C$---O$^2$&2.029&2.113\\
~Fe$^C$---O$^3$&1.920&2.136\\
~Fe$^C$---Fe$^A$&3.179&3.365\\
~Fe$^C$---Fe$^B$&3.032&3.006\\
\hline
\hline
\end{tabular}
\end{center}
\end{table}
Using the relaxed geometries, the electronic structure was studied at 
both the GGA and GGA+$U$ levels with $U=5.2$~eV. 
Local atomic charges have been calculated according to the Bader partitioning
scheme~\cite{tec-bader} using a grid-based algorithm developed by 
Henkelman et al.\cite{tec-hen-06,tec-san-07}
In our calculations we favored the Bader scheme over the Mulliken 
partitioning scheme\cite{tec-mul-55} because we found it to be more 
resistant to the difference between molecular and crystalline environments. 
The Bader charges, coupled with information from the DOS
projection of the \orb{Fe}{3}{d} orbitals, are indicative of the
oxidation state of the complex upon adsorption on the surface.
The results for the two systems are presented in the following two sections.

\subsection{Double corner sharing on the (110) surface}

After binding to the (110) surface, the Bader charge of the Fe ion in the 
complex shows a slight increase, from 1.61 to 1.68~e.
The two neighboring surface Fe ions, with which the adsorbed complex 
shares two hydroxyl ligands, show a correspondent reduction in charge 
from 1.86 to 1.82~e.
Compared with the Bader charge of 1.81~e calculated for Fe in bulk
goethite, these values do not indicate a significant transfer of
electron density typical of an oxidation process.

As far as the projected DOS around the Fermi level is concerned,
imposing an AFM ordering between the surface and the added complex
results in non-integer total spin of the system (4.8~$\mu_B$), due to
trapping of the HOMO of the complex and the previously vacant minority
spin \orb{Fe}{3}{d}-like surface states at the Fermi level.
Reversing the spin direction of the complex (i.e. choosing the
majority spin of the complex to be the same as that of the adjacent Fe
ions in the surface) results in a total spin for the system of
\mbox{4~$\mu_B$}. (The calculated total energy of this system is 0.09~eV
higher than in the previous case). 
This enables a weak interaction between the occupied minority spin 
\orb{Fe}{3}{d} orbital of the complex and the previously unoccupied 
3\textit{d} orbitals of the neighboring Fe ions in the
surface.
With a smearing width of 0.01 eV, it is possible to resolve a
HOMO (occupancy 0.90) and LUMO (occupancy 0.10), 
separated in energy by 0.025 eV.
However, the weakness of the interaction combined with its
half-metallic character lead us to believe that this is as an artifact
arising from the unphysically small band gap in the electronic
structure of goethite at the GGA level.

\begin{figure*}
\includegraphics[width=17.78cm]{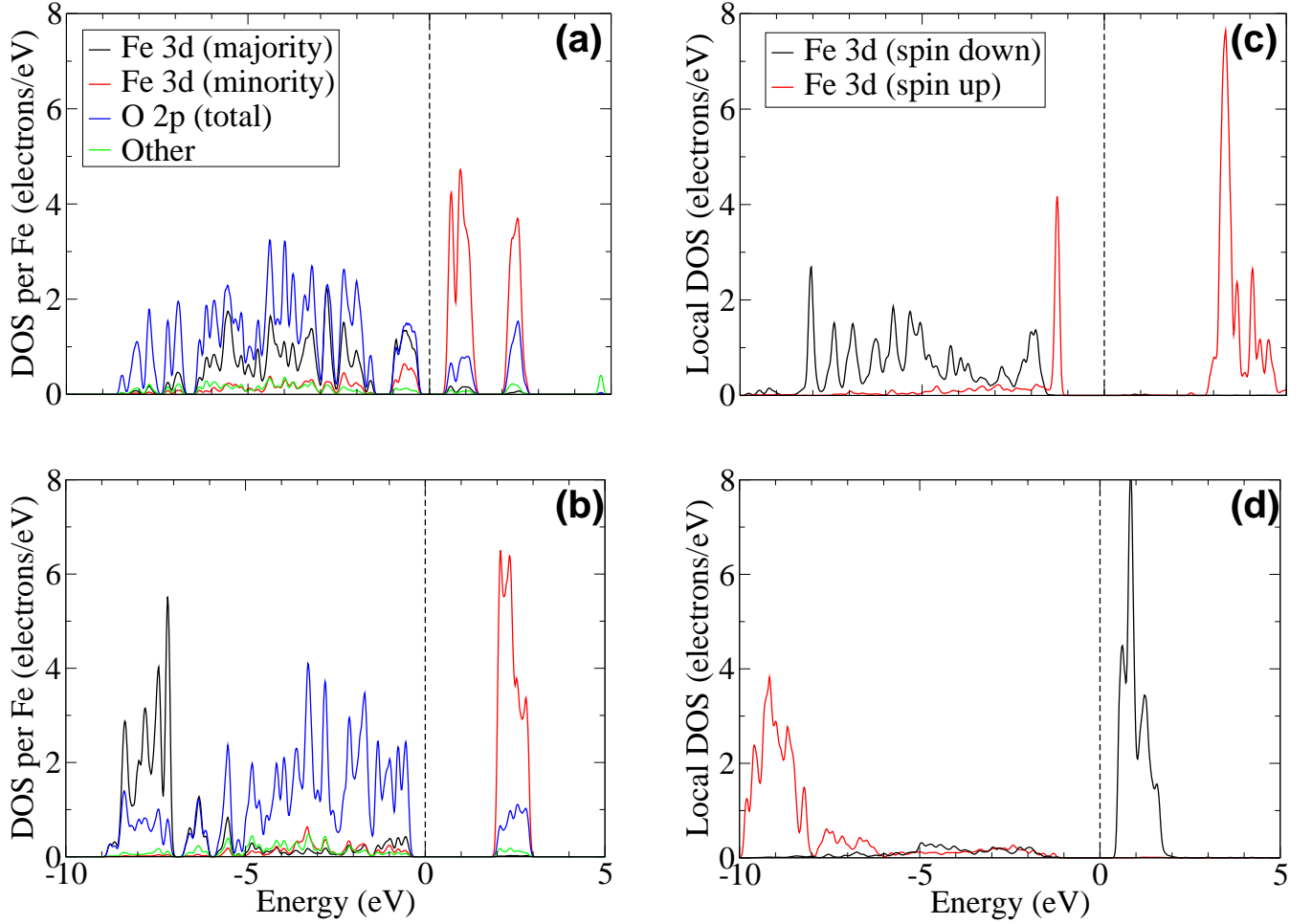}
\caption
{(Color) Calculated electronic density of states projected onto local 
Fe(3\textit{d}) orbitals: (a) bulk goethite GGA, (b) bulk 
goethite GGA+$U$, (c) complex on (110) GGA+$U$, (d) 
adjacent Fe for complex on (110) GGA+$U$.\label{fig_bulk_110_DOS}}
\end{figure*}

At the GGA+$U$ level, irrespective of the majority spin
direction of the complex, we obtain an integer total spin of
4~$\mu_B$.
In contrast to the GGA case, the Bader charge of the adsorbed Fe ion
decreases slightly, from 1.67 to 1.61~e, while the charges on
the surface Fe ions remain within 0.02 of the values for 
the bare surface.
The projected density of states on the bound complex shows a single
occupied \orb{Fe}{3}{d} minority spin orbital, characteristic of the
Fe(II) oxidation state (Fig.~\ref{fig_bulk_110_DOS}c).
Meanwhile, the minority spin \textit{d}-orbitals of the surface Fe 
ions remain safely above the Fermi level, as shown 
in fig.~\ref{fig_bulk_110_DOS}d. 
We therefore conclude that these ions remain in the Fe(III) 
oxidation state expected for bulk goethite.
%
%
%
%
%
\begin{figure*}
\includegraphics[width=17.78cm]{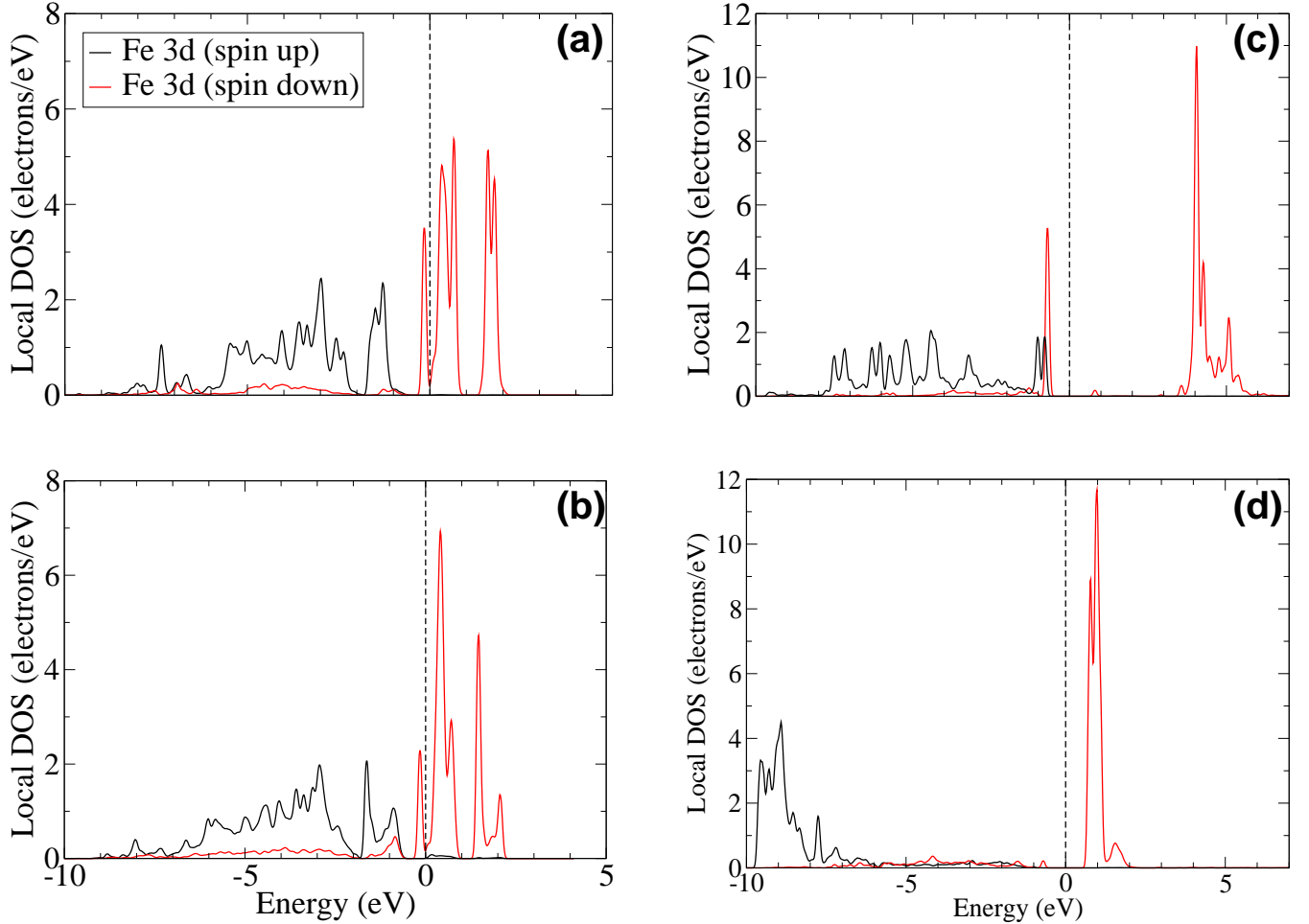}
\caption
{(Color) Calculated electronic density of states projected onto local
Fe(3\textit{d}) orbitals: (a) complex on (021) GGA, (b) adjacent Fe for complex on (021) GGA, 
(c) complex on (021) GGA+$U$, (d)
adjacent Fe for complex on (021) GGA+$U$.\label{fig_021_DOS}}
\end{figure*}

\subsection{Double edge sharing on the (021) surface}\label{autocat_021}

We now consider the double edge sharing site on the (021) surface,
starting as before with the GGA results. 
On binding to the surface, we observe an increase of the Bader 
charge on the adsorbed complex, from 1.61 to 1.66~e, and a corresponding
increase of the spin moment from 3.66 to 3.83~$\mu_B$. 
More importantly, this is accompanied by a significant decrease 
in the Bader charge of the nearest surface Fe ion from 1.89 to 1.71~e.
This value is 0.10~e lower than for an Fe ion in
bulk goethite and is thus clear evidence for partial electron transfer
from the complex to the oxide.

\begin{figure*}
\includegraphics[width=17.78cm]{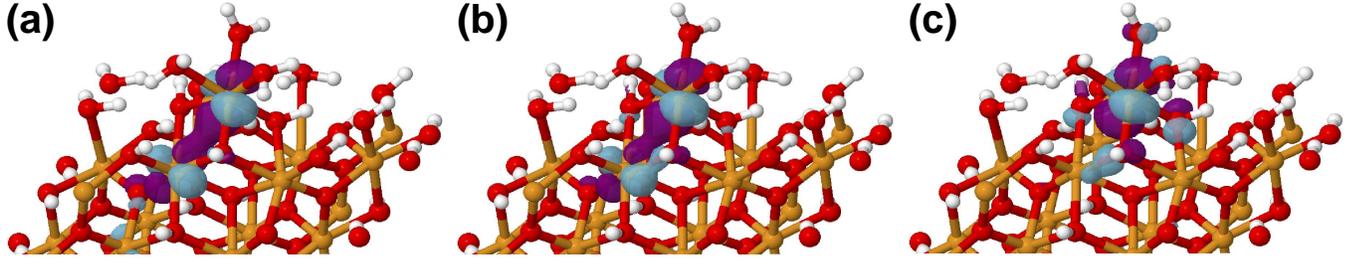}
\caption
{(Color online) The HOMO for an Fe(II) hex-aqua complex bound to the goethite (021) 
surface, calculated by the GGA+$U$ method with (a) $U=0$, (b) $U=3.0$~eV, 
(c) $U=5.2$~eV.\label{fig_021_HOMO}}
\end{figure*}

To investigate this issue further, we examine the density 
of electronic states projected onto atomic 3\textit{d}-orbitals 
on the same two Fe ions, as shown in 
figs.~\ref{fig_021_DOS}a and \ref{fig_021_DOS}b.
On each ion, the projected DOS reveals an occupied majority 
spin \textit{d}-shell and a mostly unoccupied set of minority 
spin levels, but with a single distinct minority spin peak 
0.15 eV below the Fermi level.
We previously identified such a peak as a signature of the Fe(II)
oxidation state, and its presence in the DOS of the surface Fe ion
therefore indicates partial reduction of this Fe$^{3+}$ ion
after binding of the Fe$^{2+}$ complex.
This state, which represents the HOMO of the overall system, is
localized over both ions, as depicted in fig.~\ref{fig_021_HOMO}a.
%
%
The integrated projected DOS amounts to 0.50~e on the adsorbed ion and
0.32~e on the adjacent surface ion (the remainder being attributable
mainly to \orb{O}{2}{p} orbitals).

This behavior is not, however, reproduced at the GGA+U level 
with \mbox{$U=5.2$ eV}. 
In this case, the Bader charge of the complex Fe ion is
reduced from 1.67 to 1.56~e on binding to the surface.
As before, the Bader charge of the adjacent surface Fe ion 
decreases from 1.99 in the bare surface to 1.91 with the 
adsorbed complex.
However, unlike in the pure GGA case, this does not represent a
significant reduction below the expected bulk value of 1.93.
Consistently, the projected DOS on the surface ion (fig.~\ref{fig_021_DOS}d) does 
not show an occupied minority spin 3\textit{d}-orbital, and is seen to be very
similar to the DOS previously obtained in the case of the (110)
surface (fig.~\ref{fig_bulk_110_DOS}d).
Instead, this feature is clearly evident on the adsorbed complex
(fig.~\ref{fig_021_DOS}c), with an integrated DOS of 0.75~e.
The HOMO state of the whole system at the GGA+U level is now visibly
localized only on the adsorbed ion (fig.~\ref{fig_021_HOMO}c),
confirming that the added complex retains its Fe(II) oxidation state.

To test the sensitivity of these conclusions to the value of the
Hubbard correction, we repeated the GGA+$U$ calculation with 
a lower value of \mbox{$U=3.0$ eV}. 
With this value, we obtain a result intermediate between the GGA case 
the case with $U=5.2$~eV. 
Namely, the HOMO (fig.~\ref{fig_021_HOMO}b) shows significant
interaction between minority spin \orb{Fe}{3}{d} orbitals on the
adsorbed complex and the adjacent surface ion, although the orbital
is weighted slightly more towards the complex than in the GGA ground
state (the integrated DOS amounts to 0.65 and 0.25~e for the complex
and surface ions, respectively).\cite{mixed_U}

\subsection{Discussion}

In the two previous sections we looked for possible signatures of 
spontaneous oxidation of a [Fe(H$_2$O)$_6$]$^{2+}$ complex
after adsorption on either the  (110) or (021) goethite surfaces.
On the (110) surface, we find no evidence for significant electron
transfer from the complex to the oxide at the GGA or GGA+$U$ levels.
In this case, the Hubbard term appears to correct a
presumed artefact arising from the severe underestimate 
of the band gap, which results in a semi-metallic behavior
of the system.

On the other hand, the GGA results for the (021) surface show 
a much stronger overlap between minority spin \orb{Fe}{3}{d} 
orbitals on the complex and the neighboring surface Fe ion. 
As a result, one electron is delocalized between these two ions,
which can be thought to be in a shared Fe(II)--Fe(III) oxidation 
state.
The effect of the $U$ correction in this case is to suppress such
delocalization, to an extent which we found to depend strongly on the
chosen value of $U$.
In particular, using the value of $U=5.2$~eV optimized for 
bulk goethite, the delocalization is eliminated completely, but a
value of $U=3.0$~eV still allows significant sharing of 
electron density between the complex and the surface.
This illustrates the importance of choosing the ``correct'' value
of $U$ for a given system in order to get an accurate description of
the electronic structure.

The partial electron transfer observed for the (021) but not for the
(110) surface seems to be consistent with studies of mixed valence
minerals such as magnetite, which show that the electron sharing
between neighboring ions arises from overlap of \orb{Fe}{3}{d}
orbitals in octahedral edge-sharing environments.\cite{feo-she-87}
In other words, the different behavior on the two surfaces may be 
due to the difference between the corner sharing (favored on the
(110) surface) and the edge-sharing (favored on the (021) surface) 
adsorption sites.
However, we have to note that the clear electron sharing observed at the 
GGA level is progressively reduced at the GGA+$U$ level as the Hubbard 
parameter $U$ increases.
On the basis of our calculation, we are thus led to conclude that the 
Fe(II) hex-aqua complex is not spontaneously fully oxidized on binding 
to either of the two goethite surfaces.

In light of this conclusion, the experimental results recently reported by
Wilson \textit{et al.}~\cite{surf-wil-04} on the adsorption of Fe(II)
complexes on Fe(III) oxyhydroxide surfaces are puzzling. 
As mentioned in the introduction, their M\"ossbauer spectroscopy study 
showed clear evidence for a transfer of electrons from hydrated Fe$^{2+}$ 
ions to localized sites underneath the oxide surface. 
One possible explanation is that defects or vacancies in the crystal,
not considered in our present study, might act as electron traps. 
Iron vacancies, for instance, are well known to be present in 
significant quantities even in well crystallized iron oxyhydroxides,
where they play an important role in determining magnetic 
properties.\cite{feo-ozd-96}
In particular, a large concentration of defects was found to lower
the N\'eel Temperature to 250~K, which would imply a paramagnetic
state of goethite at room temperature.\cite{Lee_2004}
Further investigation is needed to determine whether defects could
indeed enhance the ability of surrounding ions to accept electrons,
and thus provide an explanation for the apparent discrepancy between
the existing experimental finding and our simulations.

\section{Oxidation by Molecular Oxygen}
\label{sec:oxi_mol_oxy}

In the previous section we have addressed the possibility of spontaneous 
oxidation of an adsorbed Fe$^{2+}$ hex-aqua complex on the
goethite surface.
Although some electron delocalization between complex and 
surface is observed in the case of adsorption on the (021) surface, 
no clear evidence for a net electron transfer process is found 
in the absence of defects.
In this section, we investigate whether the surface, rather
than drawing electrons away from the adsorbed Fe(II) complex,
might instead increase its ability to lose electrons to an external 
oxidizing agent. 
In many cases where an iron oxyhydroxide phase grows from aqueous 
Fe(II), the net reaction taking place at the mineral surface involves 
oxidation by molecular oxygen:
\begin{equation}\label{eq_oxidation}
\mathrm{4Fe^{2+}_{(aq)}}+\mathrm{O_2}+\mathrm{6H_2O}\rightarrow
\mathrm{4FeOOH_{(s)}}+\mathrm{8H^+}\,.
\end{equation}
This equation does not represent a single step chemical process, but 
rather a complex multi-stage reaction, and a deep investigation of 
the full reaction mechanisms is outside the scope of the present work.
We thus restrict ourselves to investigating a possible first step of the 
global reaction, namely the interaction of an oxygen molecule with a 
single Fe$^{2+}$ complex, first isolated and then adsorbed onto the goethite 
(110) or (021) surfaces. 

The ground state of the oxygen molecule is a spin triplet, with one
electron in each of two degenerate $\pi$-antibonding orbitals.
Any additional electron donated from the complex to the molecule
during the oxidation will be transferred to these orbitals. 
Increasing the occupation of the antibonding orbitals will weaken the 
O--O bond, leading to an increase in the bond length and eventually,
especially via interaction with solvent water molecules, to
bond dissociation.
Here, we are looking at how the interaction differs depending
on the surface adsorption mode of the complexes, as this might give 
some insight into the catalytic role played by the mineral surface.

\subsection{Isolated Complexes}

Since Fe ions are almost universally observed to be six-fold
coordinated in aqueous solution, we assume that an oxygen molecule
will bind to an [Fe(H$_2$O)$_6$]$^{2+}$ complex  by substituting 
for one of the water ligands, forming the species 
[FeO$_2$(H$_2$O)$_5$]$^{2+}$. 
The oxygen molecule may bind to the central Fe ion either in an
end-on configuration via only one of the two oxygen atoms, or 
a side-on configuration, via both O atoms with (approximately) 
equal Fe--O lengths. 
For both configurations, we assume that the majority spin direction 
of the Fe ion and the oxygen molecule prior to the interaction are 
aligned, giving a total spin of $6/2$, so that an electron can in 
principle be donated directly from the (minority spin) HOMO of the 
complex into one of the $\pi$-antibonding orbitals of the oxygen molecule.

The structure of both possible configurations in vacuo was optimized
by means of GGA DFT calculations, using a cubic supercell of side 
length 12~\AA\ and a single k-point at the center of the Brillouin zone. 
In both cases, we observe an increase in the O--O bond length from 
the reference value of 1.23~\AA\ (obtained for an isolated oxygen molecule), 
to 1.25~\AA\ and 1.29~\AA\ for the end-on and side-on configurations, 
respectively. 
The total energy of the relaxed system is lower by 0.14~eV in the 
side-on configuration.
The Bader charge on the central Fe atom is increased by 0.10 and 0.22~e
upon binding in the end-on and side-on configuration, respectively,
consistent with a small electron transfer to the oxygen molecule.
Correspondingly, the net negative charge on the molecule increases
by 0.14 and 0.30~e in the two respective cases.
In both cases, the changes in the local spin moment of Fe are 
consistent with the changes in the charges and confirm that
the observed electron transfer involves primarily minority
spin electrons.

\begin{figure}
\includegraphics[width=8.6cm]{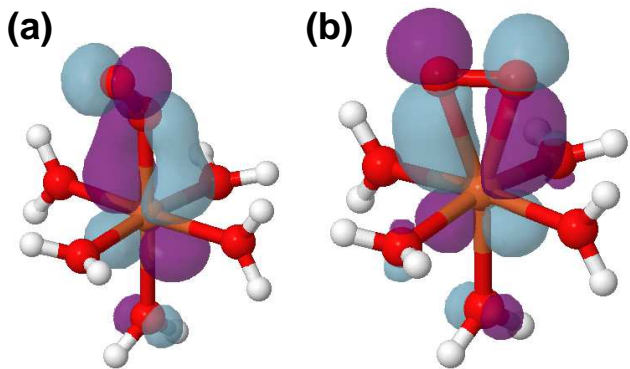}
\caption{(Color online) Highest occupied minority spin orbital of
  [FeO$_2$(H$_2$O)$_5$]$^{2+}$, from GGA calculations, 
  for end-on (a) and side-on (b) orientations of the oxygen molecule.
\label{fig_oxycomp_HOMO}}
\end{figure}

The minority spin orbitals responsible for bonding and the resulting 
electron transfer are shown in fig.~\ref{fig_oxycomp_HOMO}, showing,
as expected, strong overlap between an occupied \orb{Fe}{3}{d} and one
of the unoccupied $\pi$-antibonding molecular orbitals on the oxygen molecule.
%
%
The newly formed molecular orbital is of bonding character with respect 
to the Fe--O bonds and antibonding with respect to the O--O bond.
The formation and occupation of this orbital rather than the
original \orb{Fe}{3}{d}-dominated HOMO of the hex-aqua complex
therefore explains both the observed transfer of electron density from
the Fe ion to the oxygen molecule and the resulting weakening of
the O--O bond. 
The stronger interaction in the side-on case may be attributed 
simply to greater overlap between the iron and oxygen orbitals 
in this configuration.

In contrast with the GGA results presented above, geometry
optimization of the oxy-complexes at the GGA+$U$ level (using a value
of \mbox{$U=3.7$ eV} determined self-consistently for an Fe(II)
hex-aqua complex) causes the oxygen molecule in both configurations to
dissociate spontaneously from the complex, with the five water ligands
rearranging themselves to fill in the gap.
We note that our simulations are performed in vacuo, and that the
presence of further hydration shells may in principle influence the
stability of the oxygenated complex.
In order to obtain a reference point with which subsequent structures of
complexes adsorbed to the surface will be compared, we performed a
constrained geometry optimization of the side-on configuration,
starting from the GGA-optimized structure and allowing only the 
two oxygen atoms of the oxygen molecule to move.
In this case, the oxygen molecule remains bound, albeit very
weakly, with an average Fe--O distance of 2.34~\AA\ and an O--O 
bond length of~1.26 \AA. 
The oxygen molecule displays a net negative Bader charge of magnitude
0.12 e. 
The reduction in the interaction between the Fe ion and the oxygen
molecule resulting from the Hubbard $U$ correction is most
likely due to the lowering in energy of the minority spin \orb{Fe}{3}{d}-dominated
HOMO of the complex relative to the LUMO of the oxygen molecule. 
Indeed, increasing $U$ to 5.2~eV (which is the self-consistent value for
bulk goethite), the oxygen molecule remains electrostatically neutral and
adopts a less symmetrical position with Fe--O bond lengths of 2.59
and 2.75 \AA\ and an O--O bond length of 1.24 \AA. 

\subsection{Adsorbed Complexes}

We now turn our attention to the interaction of an oxygen molecule
with an Fe(II) aqua complex adsorbed on the (110) and (021) goethite
surfaces in the two adsorbed geometries investigated in
Section~\ref{autocat}.
As for the isolated complexes, an oxygen molecule is 
substituted for one of the free water ligands.
In light of the results of the previous section, the oxygen molecule
is placed in a side-on orientation with respect to the central Fe
ion, in order to maximize the resulting interaction.

\begin{figure}
\includegraphics[width=8.6cm]{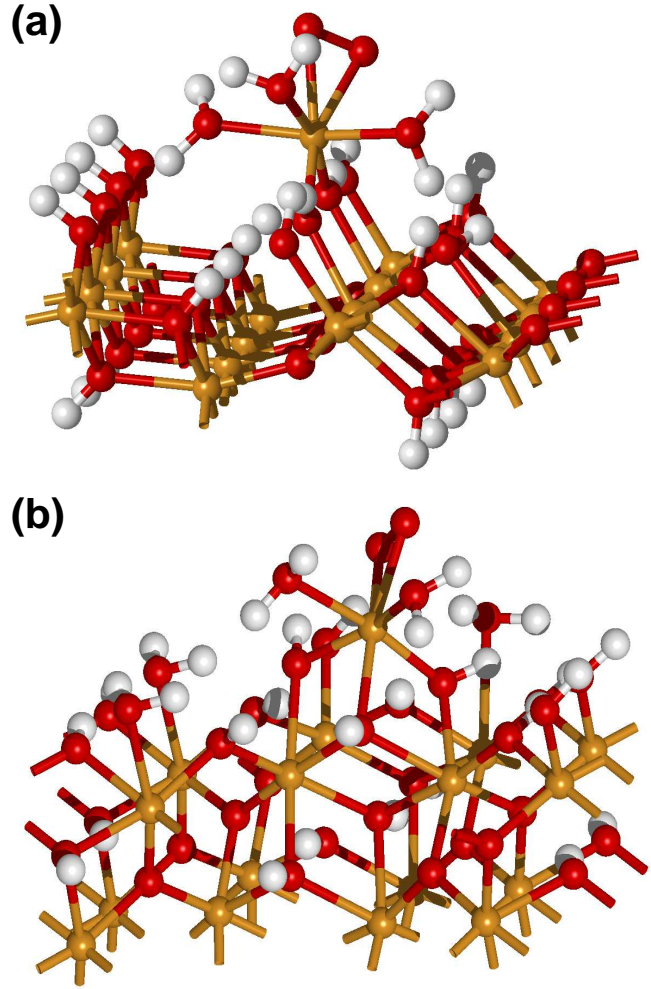}
\caption
{(Color online) The GGA relaxed structure of [FeO$_2$(H$_2$O)$_5$]$^{2+}$ bound to the
  goethite (110) surface in a double corner sharing configuration (a) and to the 
  goethite (021) surface in a double edge sharing configuration (b).
\label{fig_oxy_surface}}
\end{figure}

Both systems were fully relaxed at the GGA level, leading to the
structures shown in fig.~\ref{fig_oxy_surface}. 
%
%
GGA+$U$ calculations were also carried out, using the same value of
$U=5.2$~eV as previous calculations, starting from the optimized
GGA structures and allowing only the O atoms of the bound O$_2$ 
molecule to move according to the GGA+$U$ forces.
In all cases, the O--O bond length increases
significantly relative to the reference gas-phase value of 1.23~\AA.
At the GGA level, the relaxed O--O distance is increased by 6.8\,\% 
on either surface, compared with the increase of 4.6\,\% obtained for 
the isolated complex. 
At the GGA+$U$ level the O--O bond length increases by 3.6\,\% on 
the (110) surface and by 4.3\,\% on the (021) surface.
Notably, for the isolated complex the increase was less than 1\,\% 
at this value of $U$.
Consistently with the weakening of the O--O bond, the bound O$_2$
molecule presents a Bader charge of -0.50 or -0.52 e at the GGA level
for the (110) or (021) surfaces, respectively.
At the GGA+$U$ level, the corresponding values are -0.26 and -0.34 e.

These results give good evidence that the transfer of electron
density to the oxygen molecule and the resulting weakening of the
O--O bond are greatly enhanced for Fe(II) complexes bound to the
goethite surface with respect to free solvated complexes.
The effect appears to be slightly stronger on the (021) surface 
than on the (110) surface, especially in the GGA+$U$ calculations, 
but the difference is small compared with the overall enhancement.
In all cases the Bader analysis reveals an increase in the positive 
charge of the complex Fe ion resulting from the presence of the 
oxygen molecule.
At the GGA+$U$ level, this is the first sign of partial oxidation
of this ion in any of the systems studied. 
Interestingly, however, contrary to the charge donated to the
oxygen molecule, the increase in the positive charge on the Fe
atom is smaller for bound than for isolated complexes.
Furthermore, the increases in the charge of the Fe ion are in
all cases too small to account fully for the negative charge on the
oxygen molecule.

\begin{figure}
\includegraphics[width=8.6cm]{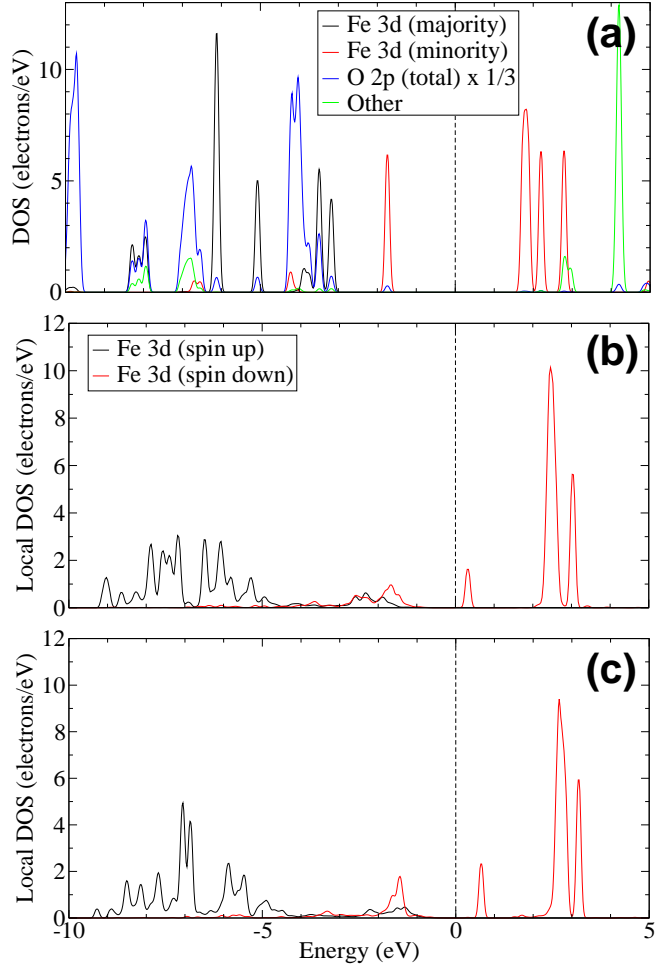}
\caption
{(Color) Calculated electronic density of states (GGA+$U$) projected onto local 
Fe(3\textit{d}) orbitals: (a) [Fe(H$_2$O)$_6$]$^{2+}$ complex, 
(b) oxy complex on (110) (c) oxy complex on (021).
\label{fig_oxycomp_surf_DOS}}
\end{figure}

To investigate this issue further, we look at the \orb{Fe}{3}{d} projected
density of states for the oxy-complex adsorbed on either surface
compared with an isolated Fe$^{2+}$ hex-aqua complex.
This is shown in Fig.~\ref{fig_oxycomp_surf_DOS} for the GGA+$U$ case
(similar results were obtained at the simple GGA level).
%
%
The DOS of the Fe(II) hex-aqua complex shows a single sharp peak in 
the minority spin density of states just below the Fermi level, representing 
the single occupied 3\textit{d} orbital that characterizes the Fe(II) oxidation
state.
The DOS of the oxygenated complexes  show two smaller peaks either 
side of the Fermi level, corresponding to bonding and antibonding
combinations with the $\pi$-antibonding orbital on the oxygen
molecule.
The occupation of only the bonding combination therefore
represents a partial oxidation of the complex Fe ion, with the extra
electron density donated to the oxygen molecule as expected. 

It is also noteworthy that, in all cases, the surface Fe ion
neighboring the oxy-complex remains in a state almost 
indistinguishable from bulk goethite. 
In the case of the (021) surface, this is significantly different 
from the behavior of this ion in the absence of a bound oxygen 
molecule (fig.~\ref{fig_021_DOS}b).
As presented in Section~4, in that case the HOMO consisted of 
minority spin \orb{Fe}{3}{d} orbitals delocalized on the complex 
and on this ion, resulting in a partially shared Fe(II)--Fe(III) 
oxidation state between the two ions.

\subsection{Discussion}

When an oxygen molecule binds to an Fe(II) aqua complex, 
electron density is transfered into one of the O--O $\pi$-antibonding
orbitals through overlap with the single occupied minority spin
\orb{Fe}{3}{d} orbital. 
The oxygen molecule becomes negatively charged, and simultaneously 
the O--O bond is weakened, rendering it vulnerable to hydrolysis.
In our simulations, the donation of charge into the oxygen 
mole\-cule is enhanced for complexes adsorbed onto either 
the (110) or the (021) surface of goethite, a result reproduced 
by both the GGA and the GGA+$U$ calculations. 
The effect is slightly stronger on the (021) surface, but the 
difference is only a small fraction of the overall effect. 
We thus conclude that the oxidation of Fe(II) by dioxygen may be 
catalyzed by adsorption onto a goethite surface, and that the 
strength of the catalytic effect is expected to vary little
between the (110) and (021) surfaces.

The catalytic effect may thus depend negligibly on the details 
of the  interaction between the complex and the surface, which 
are significantly different for the two cases, as presented
in Section~\ref{autocat}.
We propose that the oxidation enhancement may result simply from 
the higher availability of electrons in the surface environment. 
Indeed, the fact that the negative charge on the oxygen molecule 
is only partially accounted for by the increase in the positive 
charge of the Fe(II) ion indicates that the electron density lost from the
Fe(II) ion is compensated by donation of electrons from the
remaining ligands. 
This donation process is energetically unfavorable in the presence 
of electronegative water ligands only, as for the isolated complex. 
However, ligands shared between the complex and the surface can gather 
electrons from the surrounding bulk oxide, and are therefore much better 
placed to act as electron donors. 
Indeed, on oxygenation of the (021) adsorbed complex, no Bader charge
on the nearby atoms is changed by more than 0.02~e. 
This seems to confirm the idea that the additional electronic charge 
donated to the oxygen molecule is gathered from a larger area of Fe(III)  
oxide and not from any individual ion, similar to previous findings in
the case of adsorbed Sb(III)~\cite{surf-leu-06}.

\section{Conclusions}
We have performed a computational study within density functional theory 
of the stabilities of goethite surfaces and of their interaction with
Fe(II) complexes in the context of FeOOH crystal growth from aqueous solutions.
Background calculations on bulk goethite showed that structural
properties computed at the GGA level are in good agreement with 
experiment. 
The size of the band gap and the nature of the states around the 
Fermi energy are not well described at this level of theory, but 
may be corrected by use of the GGA+$U$ method.

Ab initio thermodynamics calculations performed on the (110) and (021) 
surfaces of goethite predict full hydroxylation of both surfaces when 
in contact with liquid water.
The free energy difference between the two surfaces is too small to 
account for the observed needle-like shape of goethite crystals, 
indicating that the crystal shape may be governed by kinetic factors 
rather than thermodynamic stability.
Unequal growth rates of different surfaces could result either from 
different energy barriers for the adsorption of Fe complexes from 
solution or from different oxidation rates of adsorbed Fe(II) complexes.

Oxidation of adsorbed complexes by molecular oxygen on the (021) and
(110) surfaces has been studied in Section~\ref{sec:oxi_mol_oxy}. 
In both cases we found that the underlying surface assists the transfer
of electronic charge into the dioxygen molecule.
With respect to isolated complexes in solution, bound complexes donate
up to 0.34 electrons more into the oxygen molecule, rendering the 
O---O bond increasingly susceptible to hydrolysis.
The loss of electrons from the Fe ion to the oxygen molecule is
compensated to a large degree by donation of electrons back onto the
Fe ion through the surface ligands.
This process may play a significant role in the observed autocatalytic 
growth of Fe(III) oxides from Fe(II) complexes in oxidizing solutions. 
However, the mechanism of catalysis  appears to be independent of the 
details of the interaction between the complex and the surface, occurring 
to an almost equal extent on both the (021) and (110) surfaces. 

The two surfaces show a slightly different behavior during adsorption 
of Fe(II) aqua-complexes.
In this case, some partial sharing of electronic charge has
been observed between the complex and the surface ions on
the (021) surface but not on the (110) surface.
However, this effect is limited, especially at the GGA+U
level, and our calculations do not show spontaneous oxidation
of the complex upon binding to the surface, in apparent
conflict with experimental results.\cite{surf-wil-04}
In light of this discrepancy, we propose that defects in the
oxide structure, such as Fe vacancies, may play an important role 
in assisting electron transfer from adsorbing complexes by 
trapping electrons underneath the surface.
Further investigation is needed to test the validity of 
this suggestion.

Finally, on the basis of the results presented in Section~\ref{subsec:surf_stabil} 
we propose that Fe ions from solution may adsorb more easily on the 
(021) surface than on the (110) surface.
This is due to the presence of very weakly bound water molecules
on the (021) surface, which may be displaced with virtually no energy
barrier by binding complexes at room temperature.
Therefore, given that neither spontaneous oxidation nor oxidation via
molecular oxygen appear to proceed differently on the two surfaces 
studied, the evident anisotropy of goethite crystals may be explained 
simply by different adsorption rates of additional complexes from solution.
This seems to be consistent with the fact that goethite crystals present 
approximately the same shape regardless of whether they are grown by 
precipitation from an Fe(III) solution or by oxidation of an Fe(II)
solution.\cite{surf-boi-01,surf-wei-98}

\begin{acknowledgments}

This work has been supported by the EPSRC, UK, under grant
No. GR/S61263/01.
Part of the work has been carried out within the HPC-Europa 
Project  No. RII3-CT-2003-506079 , with the support of the 
European Community Research Infrastructure Action of the FP6.
Computational resources were provided by the Cambridge HPC 
Service, UK, and by the Hochleistungsrechenzentrum Stuttgart and
by the Zentrum fŸr Informationsdienste und Hochleistungsrechnen, 
Dresden, Germany.
The work of L. C. C. is supported by the Deutschen Forschungsgemeinschaft
within the Emmy-Noether Programme.
\end{acknowledgments}

%
%


\begin{thebibliography}{54}
\bibitem{mixed_U}
Also in an additional calculation performed using an U value of 5.2 for the
Fe atoms of the goethite slab and an U value of 3.7 for the adsorbed Fe atom
(corresponding to the self-consistent values of bulk goethite and isolated complexes,
respectively) we found that the HOMO was predominantly localised on the 
adsorbed complex, with an integrated DOS of 0.77~e.
\end{thebibliography}
\end{document}